\documentclass[useAMS,usenatbib]{mn2e}
\usepackage[fleqn]{amsmath}
\usepackage{amssymb}
\usepackage{booktabs}
\usepackage{graphicx}

\newcommand{\fxf}{fxf@astro.ox.ac.uk}
\newcommand{\cwolf}{cwolf@astro.ox.ac.uk}
\newcommand{\podsi}{podsi@astro.ox.ac.uk}

\title[SN Ia time delays from high-z SNe and 
  SFH]{Constraints on SN Ia progenitor time delays from
  high-z SNe and the star-formation history}

\author[F. F\"orster, C. Wolf, 
  Ph.\ Podsiadlowski and Z. Han]{F. F\"orster$^{1}$\thanks{\fxf},
  C. Wolf$^{1}$\thanks{\cwolf},
  Ph. Podsiadlowski$^{1}$\thanks{\podsi}, Z. 
  Han$^{2}$\thanks{zhanwen@public.km.yn.cn}\smallskip
  \\$^{1}$Dept. of Physics,
  University of Oxford, Denys Wilkinson Building, Keble Road, Oxford,
  OX1 3RH, United Kingdom\\$^{2}$National Astronomical 
  Observatories/Yunnan Observatory, 
  the Chinese Academy of Sciences, P.O.Box 110, Kunming, 650011, China}

\begin{document}

\date{Accepted 2006 March 01.  Received 2006 February 28; in original form 2006 January 19}

\maketitle

\label{firstpage}

\begin{abstract}
We re-assess the question of a systematic time delay between the
formation of the progenitor and its explosion in a type Ia supernova
(SN Ia) using the \emph{Hubble} Higher-z Supernova Search sample
\citep{str04}. While the previous analysis indicated a significant
time delay, with a most likely value of 3.4~Gyr, effectively ruling
out all previously proposed progenitor models, our analysis shows that
the time-delay estimate is dominated by systematic errors, in
particular due to uncertainties in the star-formation history. We find
that none of the popular progenitor models under consideration can be
ruled out with any significant degree of confidence. The inferred time
delay is mainly determined by the peak in the assumed star-formation
history.  We show that, even with a much larger supernova sample, the
time delay distribution cannot be reliably reconstructed without 
better constraints on the star-formation history.
\end{abstract}


\section{Introduction}

Type Ia supernovae have been used extensively as standard distance indicators
and have provided the best evidence to date for an acceleration of the
Universe \citep{rie98, per99, rie04}. Future missions,
e.g. \emph{GAIA} and \emph{SNAP}, will greatly increase the number of
detected SNe Ia and significantly reduce the statistical errors in the
determination of cosmological parameters. However, the nature of the
progenitors of type Ia supernovae is still unknown and the empirically
calibrated \emph{Phillips relation} \citep{phi93} is not fully
understood physically.

Several progenitor scenarios are under discussion, but there is no
consensus due to uncertainties in the evolutionary processes
\citep{HKN96, HKN99, LV97, lan00, HP04} and the explosion mechanism
\citep{HN00, roe05, gam05}. One of the signatures of the various
scenarios is the distribution of time delays between the formation of
the progenitor systems and their explosion, which could give rise to a
significant difference between the redshift dependence of the
supernova rate (SNR) and the star-formation history (SFH).

\citet{str04}, hereafter S04, aimed to detect this difference by studying
the distribution of 25 high-z SNe Ia in the \emph{Hubble} Higher-z
Supernova Search sample \citep{rie04}. Their approach was to infer the
mean time delay of the distribution using a Bayesian analysis, which
assumed different parametrized time-delay distributions and adopted
the star-formation history (SFH) from \citet{gia04}, hereafter
G04. They concluded that mean time delays shorter than $\sim 2$~Gyr
ought to be excluded with a 95 per cent confidence level, ruling out
essentially all progenitor scenarios currently under
discussion. In a recent re-assessment of the constraints,
\citet{str04erratum} obtained a 95 per cent lower limit ranging 
from 0.2 to 1.6~Gyr for different time-delay distributions. In the
corrected best-fitting model, the 95 per cent confidence interval
ranged from 1 to 4.4~Gyr with the most likely value at 3.4~Gyr.

Unlike core collapse SNe (SNe II, Ib/c) that originate from massive
progenitors with relatively short main-sequence (MS) lifetimes ($\sim
3-20$~Myr), SNe Ia are believed to be thermonuclear explosions of
white dwarf stars (WDs) whose progenitors have MS lifetimes ranging
from $\sim 30$~Myr to several billion years. This implies a minimum
time delay for SNe Ia of the order of $\sim 30$~Myr.

Most of the SN Ia progenitor scenarios that have been proposed involve
mass transfer on to a CO WD in a binary system, either through the
expansion and Roche lobe overflow of an evolved companion
(\emph{single degenerate [SD] scenarios}) or through the slow release
of gravitational waves, orbital shrinking, Roche lobe overflow and
merging of a compact double WD system (\emph{double-degenerate [DD]
scenarios}). Both scenarios have associated time-delay distributions
that have been estimated with binary population synthesis codes (BPS),
where the properties of binary systems are followed from their birth
up to the explosion stage through the many different evolutionary
paths. Independently of the particular treatment of the binary
interactions, the resulting time-delay distributions differ
considerably since their characteristic time-scales have different
origins.

The SD scenario is controlled by the process of mass accretion, which
has to occur at just the correct critical rate in order to allow the
growth of the mass of the companion WD up to the Chandrasekhar limit
\citep{nom91}. The dominant evolutionary path seems to occur via the
accretion of matter on to a CO WD from a slightly evolved MS star, the
CO WD + MS -- SD scenario \citep{vH92, rap94, LV97, lan00, HP04}. In
this channel, the accretion rate is determined mainly by the mass of
the donor star, which must lie in a narrow range in order to satisfy
the required accretion-rate constraints. As a consequence, the
distribution of MS lifetimes and the time-delay distribution of the
channel are relatively narrow, peaking at $\sim 670$~Myr and rapidly
becoming negligible after $\sim 1.5$~Gyr.

Although recent simulations have suggested that other evolutionary
paths within the SD framework are of minor importance
\citep{HP04}, it is quite possible that their contribution has been
underestimated. This is particularly important for the CO WD + RG --
SD scenario, where a red-giant (RG) star accretes matter on to a CO WD
star \citep{HKN96}; in this channel the time-delay distribution 
extends up to several Gyr.

The DD scenario \citep{IT84, web84}, in contrast, is controlled by the
time that it takes for the binary system to coalesce, which depends
roughly on the fourth power of the separation of the double-degenerate
system \citep{sha83}. As a result, the time-delay distribution can be
described by a low time-delay cutoff ($\sim 30-100$~Myr) and an
approximately power-law decline up to the age of the Universe. The
lower time-delay cutoff can be explained by the time required to form
the most massive degenerate systems with the shortest MS lifetimes,
whereas the power-law tail can be explained by the power-law relation
between coalescence time and separation of the double-degenerate
systems.

However, the expected accretion rates in the DD scenario are a
problem: they are so high that present calculations suggests that this
leads to accretion-induced collapse (AIC) and the formation of a
compact object rather than a thermonuclear explosion \citep{NI85,
SN85, SN98, TWT94, nom91}.

Therefore, the currently generally most favoured progenitor scenario
is the SD scenario. Because the seemingly dominant evolutionary path
of this channel would need to be discarded if the mean time-delay were
found to be higher than 2~Gyr, it is important to confirm the
significance of the S04 results.

In this work we have studied the SN Ia time-delay distribution using
the sample of S04 and the same basic analysis, but introducing
alternative SFHs found in the literature, avoiding binning effects as
much as possible and using a Goodness of Fit (GoF) test that is
generally recommended for small samples. We discuss the data and
analysis in Section \ref{sec:analysis}, show the results and
Monte Carlo simulations in Sections \ref{sec:results} and
\ref{sec:Monte Carlo} and discuss their significance in Sections
\ref{sec:discussion} and
\ref{sec:conclusions}. Throughout his paper, we adopted a value for the
Hubble constant of $H_{\rm 0} = 70$ km s$^{-1}$ Mpc$^{-1}$, present
ratios of matter, curvature and dark energy density over the critical
density of $\Omega_M = 0.3$, $\Omega_K = 0$ and $\Omega_{\Lambda} = 0.7$,
respectively, and a `dark energy' pressure over density ratio ('equation
of state') of $w =-1$.

\section{Data and Analysis} \label{sec:analysis}

The analysis is based on the \emph{Hubble} Higher-z Supernova Search
sample \citep{rie04, dah04, str04}, which contains 25 SNe Ia found in
the GOODS field (13 in the Hubble Deep Field North, HDFN, and 12 in
the Chandra Deep Field South, CDFS) in the redshift range 0.21 to
1.55. The SNe were discovered in four difference images that were
produced by observing both fields five times in intervals of
approximately one month, comparable to the typical duration of
the main SN Ia light curve peak.

To infer the underlying time-delay distribution of SNe Ia, S04
compared the observed redshift distribution in the sample with a
parametrized predicted distribution, derived from the G04 SFH
convolved with three alternative time-delay distributions.

Each distribution was parametrized by its mean time delay, which was
recovered using a Bayesian analysis. Among these, the distribution
that best fit the data was a `narrow Gaussian', which after being
corrected was centred on 3.4~Gyr with a FWHM of $\sim 1.5$~Gyr. The 95
per cent confidence interval for the mean time delay ranged from 1.0 to
4.4~Gyr. The alternatives `wide Gaussian' and e--folding distributions
had a mean time delay above 0.2 and 1.6~Gyr, respectively, with more
than 95 per cent confidence.

Only the shapes of the distributions are compared, i.e. the analysis
is scale--free, and the associated efficiencies of SNe per unit formed
mass are calculated later by normalising the models to the SN numbers
and are not used to constrain the models. This means that the sample
must ideally span a redshift range that includes both the rising and
declining parts of the SNR, i.e. where the SNR is not approximately
linear. A recent study \citep[see][Fig.~6--8]{BT05} could not fully
exploit information on the SN redshift distribution because their
sample did not reach to a sufficiently high redshift ($z>1$), as the
authors indicate in the text. A similar situation is found in the work
of \citet{GYM04}.

Thus, because the \emph{Hubble} Higher-z Supernova Search sample is
the deepest SN sample available, it is the most suitable one for
constraining the time-delay distribution of SNe Ia.

However, the formal errors quoted in S04 reflect only the limited size
of the sample and not other systematic uncertainties, such as
those associated with the SFH.

In the following Sections \ref{sec:SNR} to \ref{sec:tc} we introduce
the formalism that gives the SNR, the number of detected SNe per unit
redshift and the control times used in the derivations. In Section
\ref{sec:timedist} we discuss alternative time-delay distributions
and in Section \ref{sec:SFH} alternative SFHs. The Bayesian
analysis is described in Section \ref{sec:Bayes} and further
modifications concerning binning effects and the GoF test are
discussed in Section \ref{sec:newanalysis}.

\subsection{The SN Ia rate}  \label{sec:SNR}

The rate of SNe Ia per unit time per unit co-moving volume ($SNR_{\rm
Ia}$) is given by the star-formation rate per unit time per unit
co-moving volume ($SFR$) convolved with the normalised distribution of
explosions per unit time of the progenitor channel (the time-delay
distribution, $\phi$), and multiplied by the number of SNe per unit
formed mass (the efficiency, $\nu$). We assume that neither $\nu$ nor
$\phi$ evolve with redshift:
\begin{align}
  SNR_{\rm Ia}(z) = \nu \int_{\rm t(z_R)}^{t} SFR(t')\ \phi(t - t',
  \tau) \ dt' \label{eq:SNR},
\end{align}
where $t = t(z)$, $\tau$ is some characteristic time-scale defined in
Section ~\ref{sec:timedist} and $z_R$ is the redshift associated with
the time when the first stars formed, approximately the epoch of
reionisation. We assumed $z_R =10$, as in S04.

\subsection{Distribution of detected SNe}  \label{sec:nIa}
The number of detected SNe Ia per unit redshift interval ($n_{\rm
Ia}$) is given by the multiplication of the rate of SNe Ia per unit
time per unit co-moving volume ($SNR_{\rm Ia}$), a time dilation
factor, $(1+z)^{-1}$, the control time of the survey ($t_{\rm c}$) and
the volume per unit redshift being surveyed, $\displaystyle{
\frac{dV}{dz d\omega} \Delta \omega}$:
\begin{align}
 n_{\rm Ia}(z) = \frac{SNR_{\rm Ia}(z)}{1+z} \ t_{\rm c}(z) \
\frac{\mathrm{d}V(z)}{\mathrm{d}z \mathrm{d}\omega} \ \Delta\omega \label{eq:nIa},
\end{align}
where in our cosmology the volume derivative formula simplifies to:
\begin{align}
  \frac{\mathrm{d}V}{\mathrm{d}z \mathrm{d}\omega} = d_C^2 \frac{\mathrm{d}(d_C)}{\mathrm{d}z}\text{,  where}
\end{align}
\begin{align}
  d_C = c H_0^{-1} \int_0^z \mathrm{d}u \left[ (1+u)^3 \Omega_M +
    \Omega_\Lambda \right] ^{-1/2}
  \label{eq:dV1},
\end{align}
and hence,
\begin{align}
  \frac{\mathrm{d}V(z)}{\mathrm{d}z \mathrm{d}\omega} = c H_0^{-1} \left[ (1+z)^3
    \Omega_M + \Omega_\Lambda \right] ^{-1/2} d_C^2(z)
  \label{eq:dV2}.
\end{align}

\begin{figure}
\centering
\includegraphics[clip=true,width=0.8\hsize, keepaspectratio]{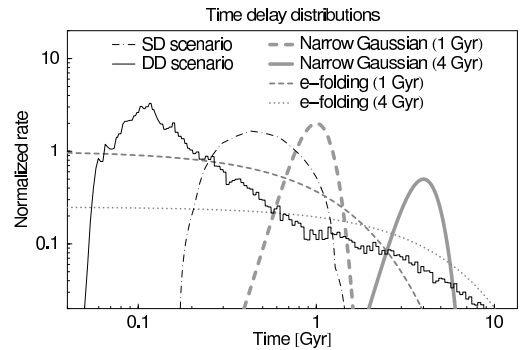}
\caption{Theoretical time delay distributions \citep{HP04} compared to
  parametrized time delay distributions used in the analysis. The
  best-fitting model in S04 corresponds to the `narrow Gaussian'
  distribution with a mean time delay of 4 Gyr}
\label{fig:loglog-TheoryDelay}
\end{figure}

\subsection{The control time} \label{sec:tc}
The control time can be understood as the total observing time
multiplied by the probability of detecting a SN at a given
redshift. We used the same values as S04, that were calculated taking
into account the expected extinction, spectra, light curve shapes and
peak magnitude dispersion of SNe Ia, the way each field was revisited,
and the efficiency of the detection algorithm (but see Section
\ref{sec:binning}). The control times are defined by

\begin{align}
t_{\rm c}(z) = \iiint P(t |
M_\lambda, A_\lambda, z) \ P(M_\lambda) \ P(A_\lambda) \ dM_\lambda \
\mathrm{d}A_\lambda \ \mathrm{d}t,
\end{align}
where $P(t | M_\lambda, A_\lambda, z)$ is the probability of detecting
a new SN at time t, given its rest-frame luminosity and its host
galaxy extinction and redshift. It depends on the assumed spectra
through K--corrections, the sensitivity of the survey and the
efficiency of the detection algorithm.

$P(M_\lambda)$ is the probability of having a given SN rest-frame
luminosity. It was estimated based on the characteristic relation
between peak luminosity and light curve shape of SNe, and the observed
dispersion of SN Ia peak luminosities.

$P(A_\lambda)$ is the probability of having a given host galaxy
extinction at the given rest-frame wavelength. It was assumed to be
proportional to $e^{- A_\lambda}$.

For more details see the original discussion in S04.

\begin{figure}
\centering
\includegraphics[clip=true,width=0.8\hsize, keepaspectratio]{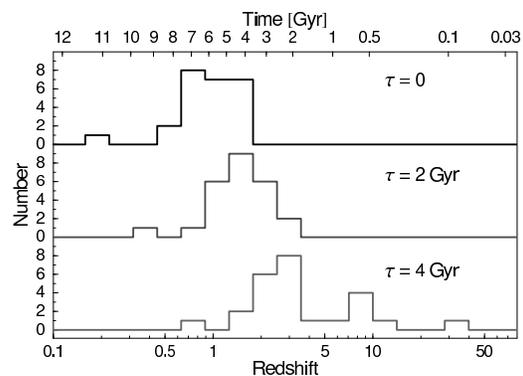}
\caption{Histogram of progenitor formation redshifts for the 25 SNe in
the \emph{Hubble} Higher-z Supernova Search sample assuming unique
time delays. Note that some SN progenitors must have originated at
extremely high redshifts if the time delays were always high.}
\label{fig:SNeHisto}
\end{figure}

\subsection{Time delay distributions} \label{sec:timedist}

All the time delay distributions were parametrized by their mean time
delays, $\tau$. S04 used an exponential distribution and two families
of Gaussian distributions whose width scale with the mean time
delay. The e-folding distributions are given by:

\begin{align}
\phi(t,\tau) = \frac{e^{-t/\tau}}{\tau}.
\end{align}
The two alternative Gaussian distributions are grouped into the
families of `narrow' ($\sigma_{\tau} = 0.2 \tau$) and `wide'
($\sigma_{\tau} = 0.5 \tau$) distributions, of the form:

\begin{align}
\phi(t,\tau) = \frac{1}{\sqrt{2 \pi \sigma_{\tau}^2}} e^{ -\tfrac{(t - \tau)^2}{2
    \sigma_{\tau}^2} }.
\end{align}
The previous distributions are defined only for positive
values. However, negative time delays must be allowed in order to
avoid statistical bias in a small sample and to get confidence
intervals that do not artificially discard short time
delays. Moreover, a preference for negative time delays would signal
SFHs that peak too late in time. Thus, we considered a fourth time
delay distribution, a Gaussian distribution with fixed width ($\sigma
= 0.5$~Gyr) that allows either for positive or negative time delays:

\begin{align}
\phi(t,\tau) = \frac{1}{\sqrt{2 \pi \sigma^2} }
 e^{ -\tfrac{ (t-\tau)^2 }{ 2 \sigma^2 }}.
\end{align}
We also added a log-normal distribution, which is associated with
processes where the source of uncertainty has multiplicative effects
rather than additive ones, as is the case for Gaussian
distributions. The best-fitting models to the theoretical time delays
were in most cases log-normal distributions, whose width $\sigma$, in
units of $\rm{log}(t)$, was kept fixed and determined by the
best-fitting model of the theoretical time delays:

\begin{align}
\phi(t,\tau) = \frac{1} {\sqrt{2 \pi \rm{ln}(\sigma)^2}} e^{-
  \tfrac{\rm{ln}(t/\tau)^2} {2\ \rm{ln}(\sigma)^2} } \frac{1}{t}.
\end{align}

The theoretical time delay distributions from \citet{HP04} were also
examined with a GoF test. They were produced assuming either the CO WD
+ MS -- SD scenario or the DD scenario with different binary evolution
parameters. In Fig.~\ref{fig:loglog-TheoryDelay} the theoretical time
delay distributions of the SD and DD scenario together with two time
delay distributions with different mean time delays are plotted,
including the best-fitting model from S04.

\subsection{The Star-Formation History (SFH)} \label{sec:SFH}

Because we consider alternative prescriptions for the SFH that have
incomplete redshift information, further complications arise. The
ideal redshift coverage of the SFH should range from zero to the
redshift of the first star formation, farther than the highest-redshift 
object presently known in the Universe. This is a consequence
of the redshift range of the detected SNe and the long time delays
that have to be considered. If high-redshift SNe only exploded after
long time delays, their progenitors would need to form at redshifts up
to $\sim 30$, as Fig.~\ref{fig:SNeHisto} shows.

If the determination of the SFH did not cover the required redshift
range, we used as an approximation either a power-law extrapolation in
time or a scaled version of the optical--UV derivation. We found that
the method is not very sensitive to this approximation when the
position of the peak of the SFH is well constrained, since it is the
difference between the peaks of the SFH and the SNR that mainly
constrains the best-fitting models. The alternative prescriptions of
the SFH we have used are the following:

\begin{itemize}
\item The SFH by \citet{gia04}, G04. We have used continuous
  approximations for the extinction corrected and not corrected models,
  inferred from deep optical--UV observations of galaxies in the GOODS
  field. Both versions differ by a factor of $\sim 3$ in the redshift
  range of interest. The continuous approximations are the ones used
  in S04, which peak at $z \sim 2.7$ in the extinction corrected model
  (M1) and at $z \sim 1.8$ in the model that is not corrected for
  extinction (M2).
\item The best-fitting model of \citet{CE01}, hereafter CE01. It was
  derived from the integrated cosmic infrared background (CIRB) and
  covers the redshift range from 0 to 4.5. At $\rm{z > 4.5}$ we tried
  a power-law extrapolation in time or a scaled version of G04 at
  $\rm{z > 3}$. Mainly because the peak of the SFH occurs very late in
  time, at $z \sim 0.8$, we found that the inferred time delays are
  not very sensitive to this approximation. However, a constant SFH
  model is within the error bars at high--z.
\item The SFH from \citet{hea04}, hereafter H04, inferred from the
  `fossil record' of stellar populations in the \emph{Sloan Digital
  Sky Survey} (SDSS). We interpolated a Spline function to the binned
  SFH, which peaks at $z \sim 0.4$, in order to obtain a smooth
  approximation. It is not usually recommended to approximate data in
  this way, but we think that our approximation preserves the general
  differences between this SFH and the alternative prescriptions. We
  also tried a scaled version of G04 at $\rm{z > 3}$, or a constant
  star-formation history for all times, since this SFH is very flat at
  $z \gtrsim 1$ and peaks very late in time with respect to the
  detected SNe. Both alternatives gave very similar results to the fit
  to the original binned data.
\end{itemize}
 
Additionally, we considered one of the most recent determinations of
the SFH using \emph{Spitzer} data, presented by \citep{PG05}. However,
both its limited redshift coverage and its dependence on the assumed
galaxy luminosity function makes the high-redshift extrapolation
ambiguous. For this reason, we did not try a continuous approximation
of this SFH in the analysis, although it must be considered a reliable
result. The variance between its different versions only demonstrates
the persisting uncertainties in our knowledge of the SFH.

\begin{figure*}
\centering
\includegraphics[clip=true,width=0.6\hsize, keepaspectratio]{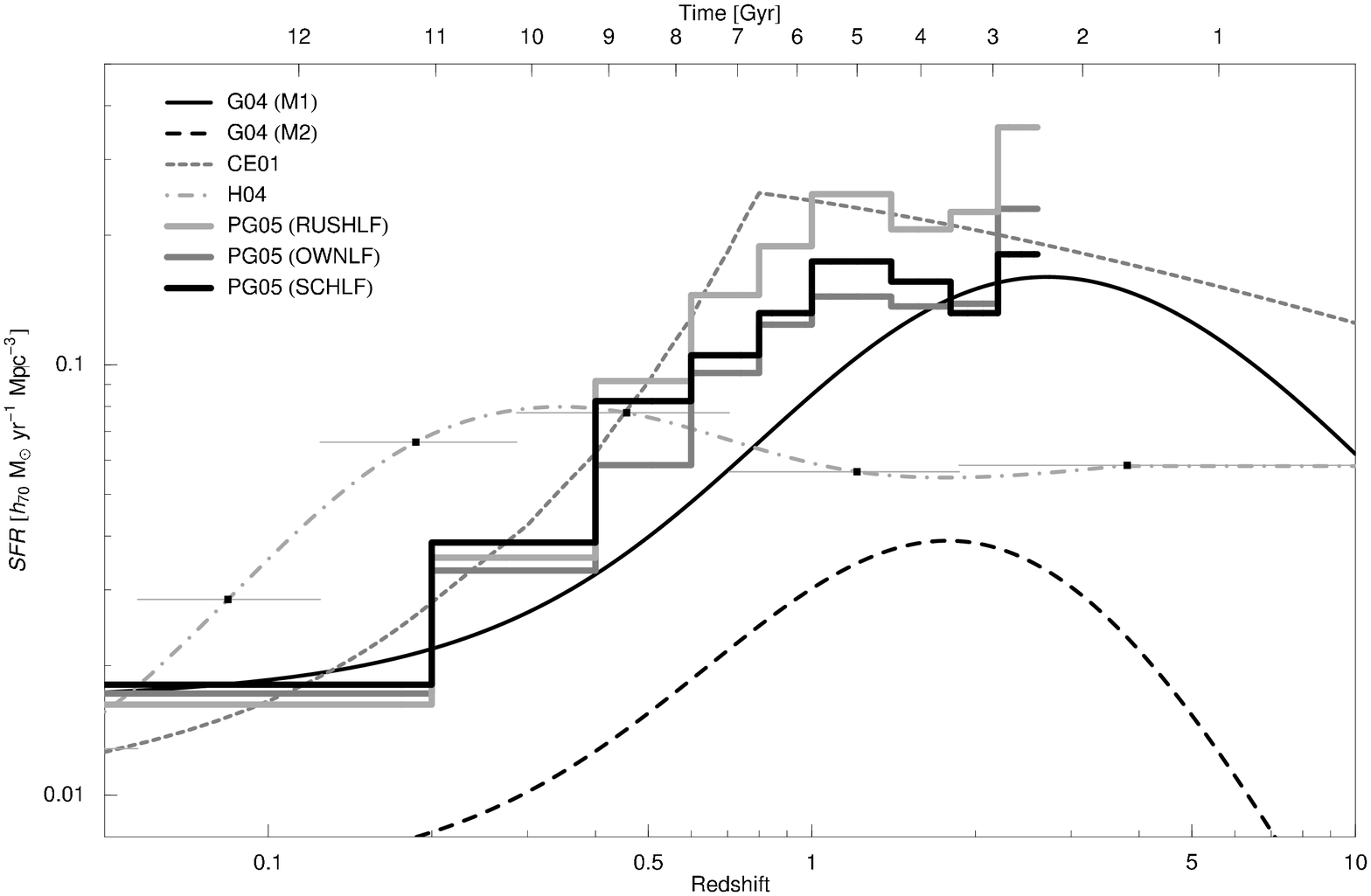}
\caption{The alternative SFHs that better reflect the diversity of
  results obtained in the analysis. For G04 \citep{gia04}, M1 is the
  extinction corrected model and M2 is not corrected. For CE01
  \citep{CE01}, we used a power-law extrapolation in time for $z>4.5$,
  but also a scaled version of G04 (not shown here), which gave very
  similar results. For H04 \citep{hea04}, horizontal error bars
  represent bin sizes and vertical error bars (very small), bootstrap
  errors. In our calculations we chose to represent it by a smooth
  Spline interpolation. The PG05 \citep{PG05} SFHs are determined
  combining optical--UV data with \emph{Spitzer} observations in the
  GOODS field, assuming different galaxy luminosity functions. The SNe
  in the \emph{Hubble} Higher-z Supernova Search sample are in the
  redshift range from $z = 0.21$ to $z = 1.55$.}
\label{fig:loglog-SFH}
\end{figure*}

In Fig.~\ref{fig:loglog-SFH} we show the four continuous
approximations of the SFH and the binned SFH from PG05. The continuous
approximations are based on the extinction corrected and not corrected
SFH from G04, the best-fitting model from CE01 with a power-law
extrapolation in time at high redshift and a continuous approximation
of the best-fitting model from H04, which is constant at high
redshift.  It is apparent that there is a range of SFHs in the
literature that do not agree and, importantly, peak at very different
times. Thus, it is important to understand the systematic errors
associated with this uncertainty. For a recent estimation of the
uncertainties on the SFH see also Fig.~2 and 4 from \citet{HB06}.

\subsection{The Bayesian probability} \label{sec:Bayes}

Using Bayes theorem with a uniform prior, the probability of a mean
time delay $\tau$ with a time-delay distribution $\phi(t, \tau)$ and a
SFH in the form of $SFR(t')$, given the set of SN redshifts, $\{ z_i
\}$, is proportional to:

\begin{align}
P( SFR(t'), \phi(t, \tau), \tau \vert \{z_i\}) \varpropto P( \{z_i\} \vert
SFR(t'), \phi(t, \tau), \tau).
\end{align}
Thus, it is proportional to the probability of the particular SN
redshift distribution:

\begin{align}
P( \{z_i\} \vert SFR(t'), \phi(t, \tau), \tau) \varpropto \prod_{i=1}^{25}
n_{\rm Ia}(z, \tau),
\end{align}
where $n_{\rm Ia}(z,\tau)$ has been normalised for every $\tau$
because the analysis is scale free. It depends on $\tau$ through the
time-delay distributions of Section \ref{sec:timedist} and equations
\ref{eq:SNR} and \ref{eq:nIa}.

Hence, for a given combination of SFH, time-delay distribution and
mean time delay, the predicted number of SNe per unit redshift can be
expressed as a probability distribution in redshift. Subsequently, the
probability of the set of SNe can be calculated for every $\tau$.

\begin{figure*}
\centering
\vbox{
\includegraphics[clip=true,width=0.3\hsize, keepaspectratio]{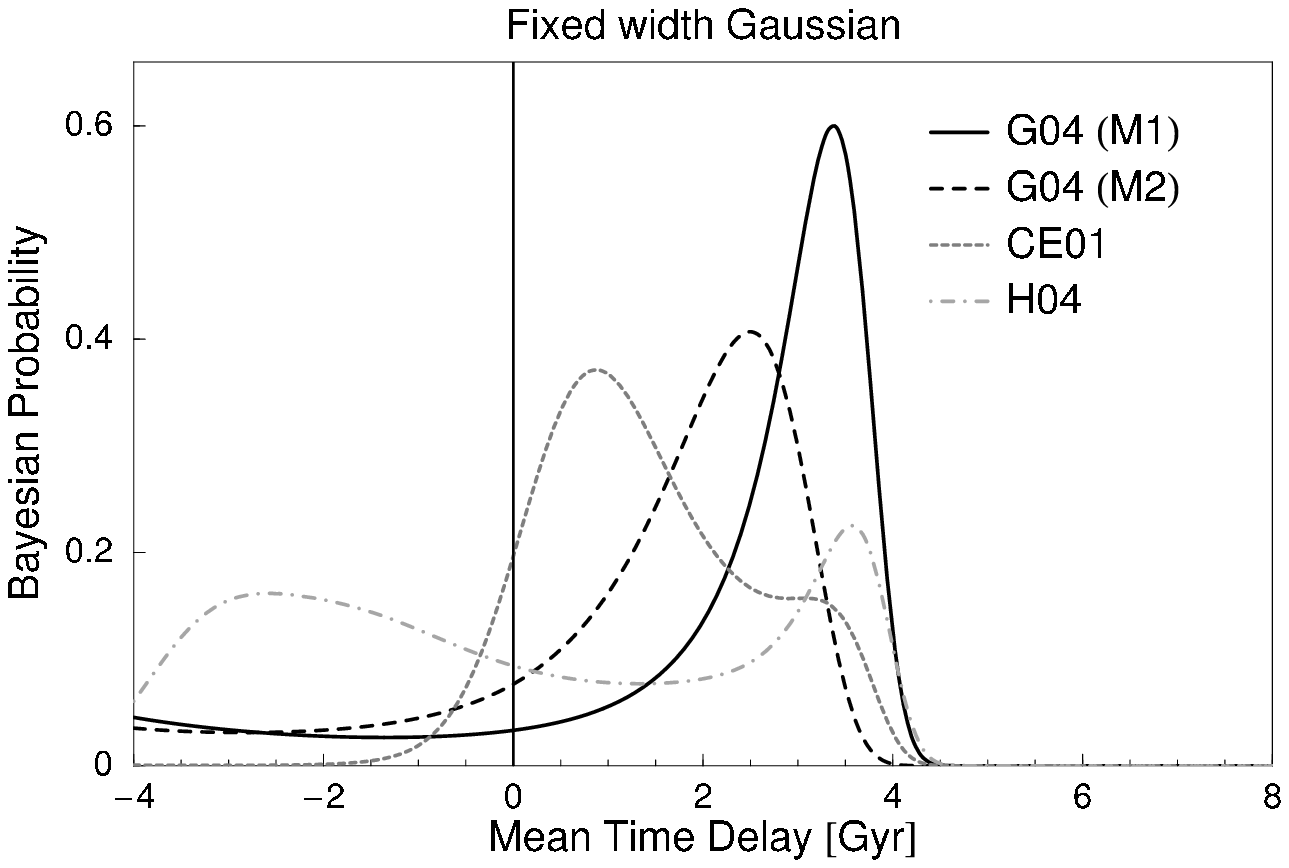}
\includegraphics[clip=true,width=0.3\hsize,, keepaspectratio]{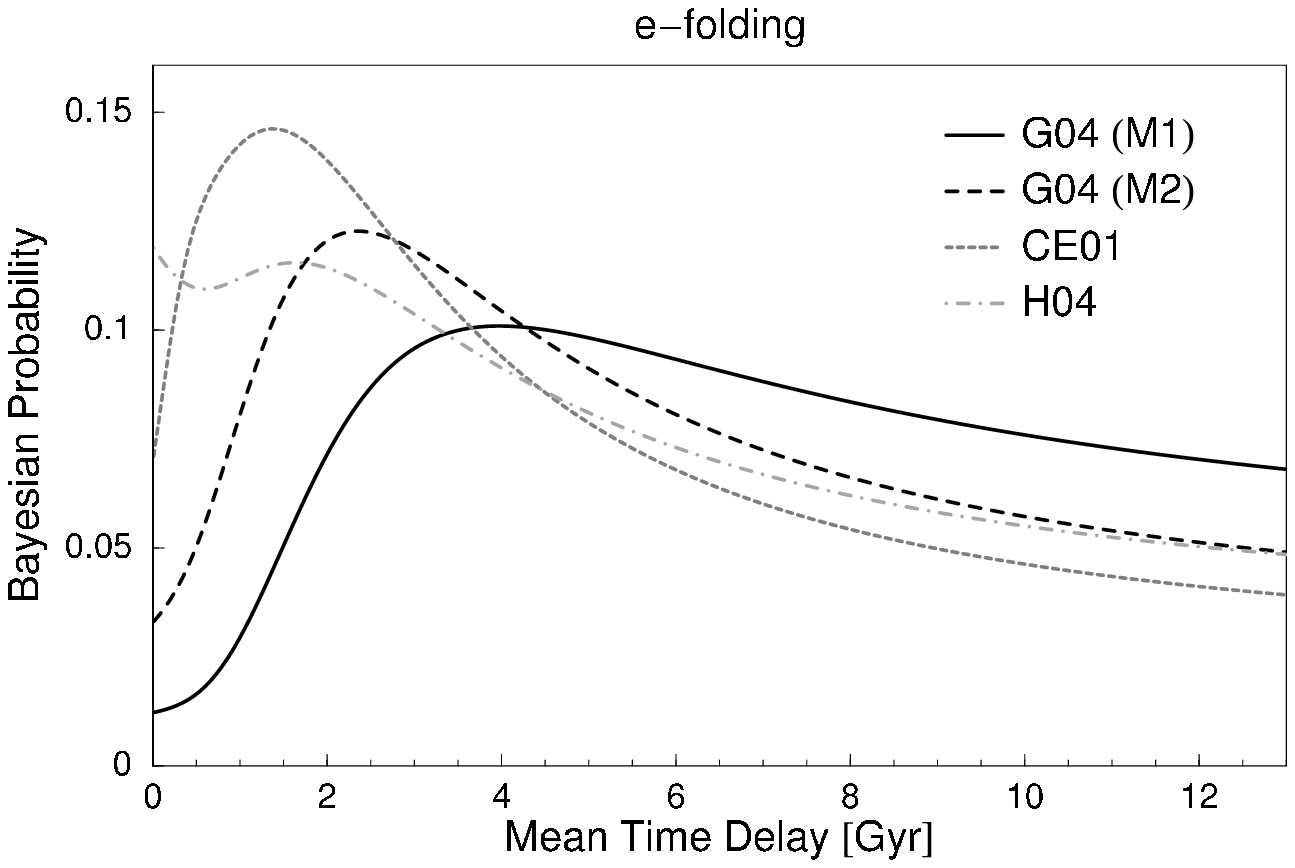}}
\vbox{
\includegraphics[clip=true,width=0.3\hsize, keepaspectratio]{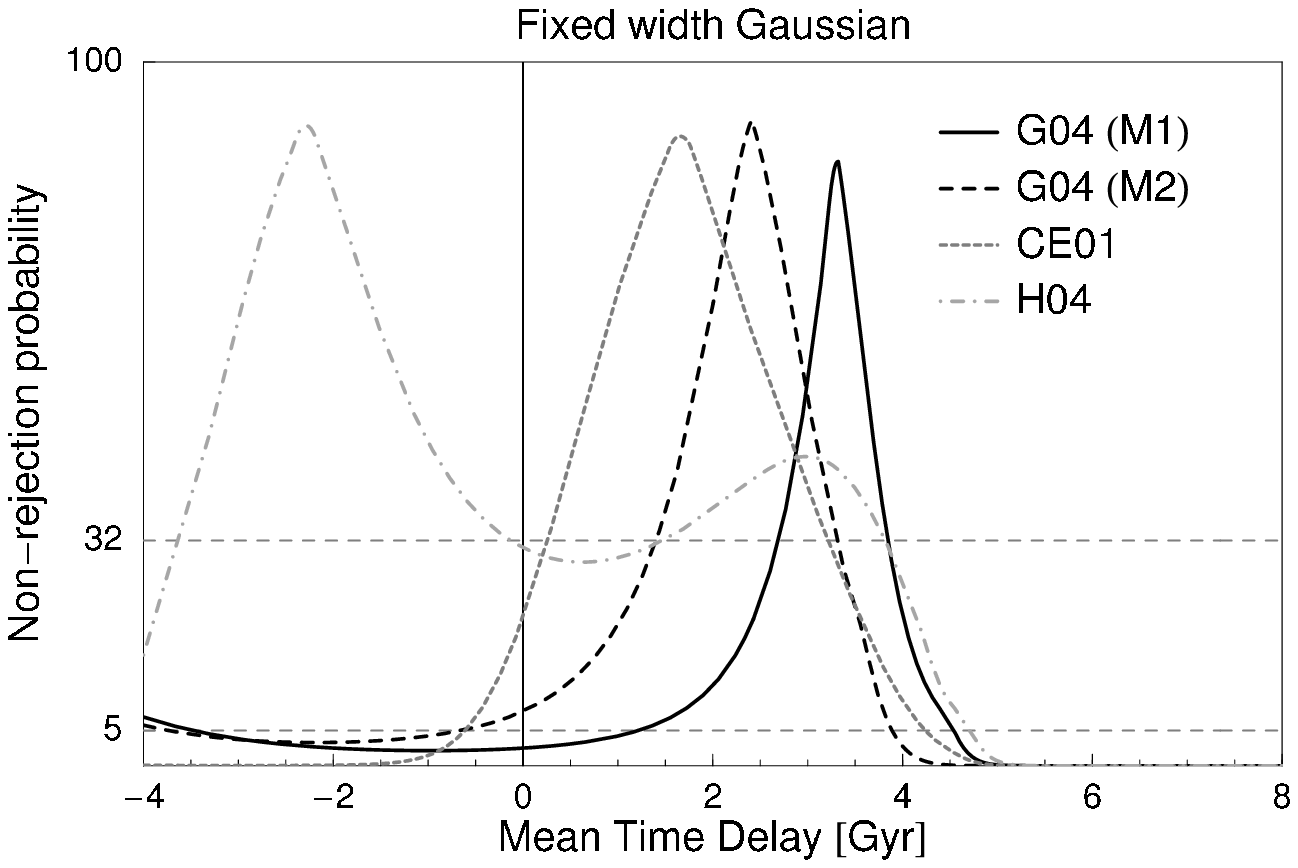}
\includegraphics[clip=true,width=0.3\hsize, keepaspectratio]{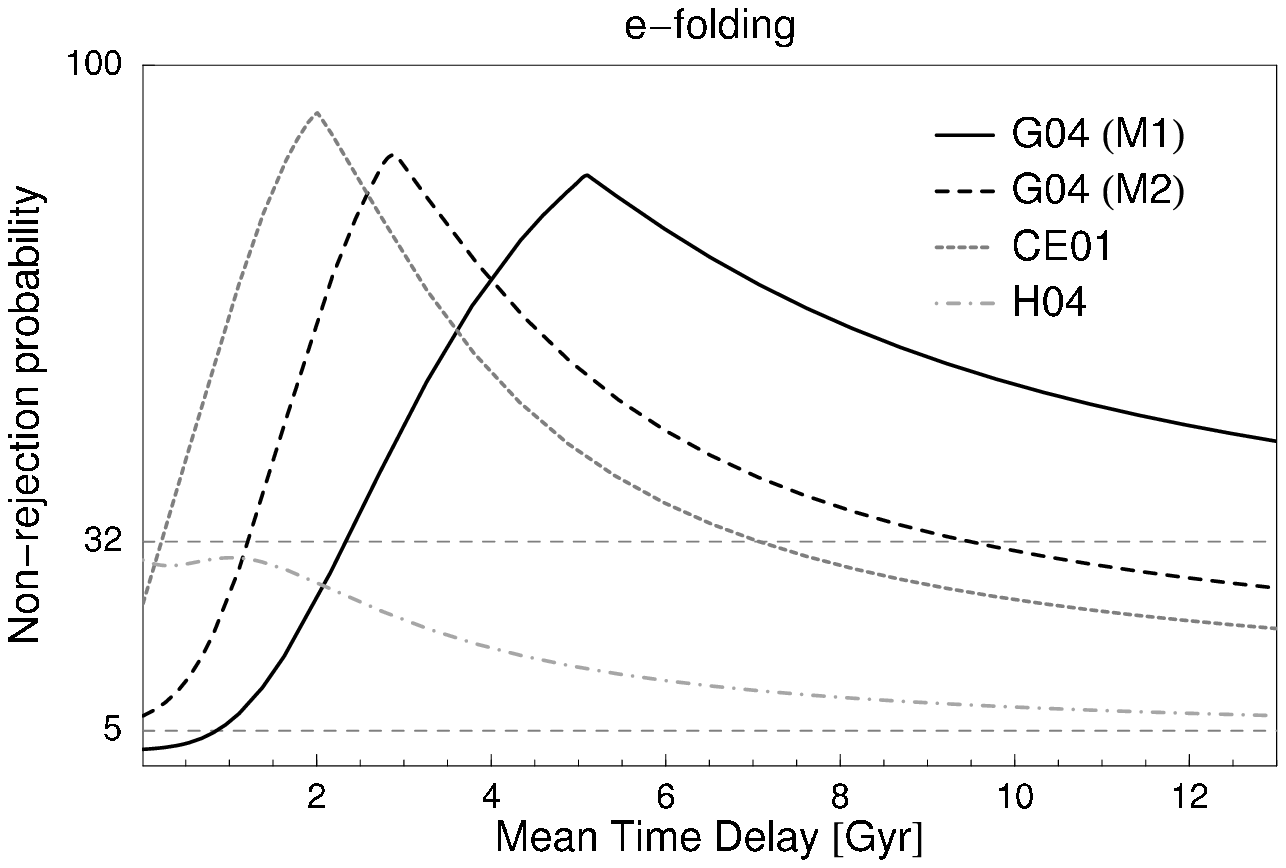}}
\caption{Bayesian probabilities and KS test non-rejection
  probabilities for the case of the `fixed width Gaussian' and
  e-folding time delay distributions.  The probabilities associated
  with the `narrow Gaussian' distribution are very similar to those
  associated with the `fixed width Gaussian' distribution in the
  positive time-delay region.  Note the relation between the most
  probable and best-fitting values with the peak of the SFHs in
  Fig.~\ref{fig:loglog-SFH}, i.e. later peaked SFHs give shorter time
  delays. The apparent preference for negative time delays in the case
  of the SFH from H04 may be an indication that this SFH peaks too
  late in time. See Section~\ref{sec:results} for more details.}
\label{fig:probs}
\end{figure*}

\subsection{New analysis}  \label{sec:newanalysis}

The main differences between the S04 calculations and this work are a
result of considering a range of alternative SFHs. Further
differences are as follows:

\subsubsection{Redshift binning effects}  \label{sec:binning}

The main advantage of using the observed redshift distribution of SNe
Ia instead of the corresponding SNR \citep{GYM04} is that the analysis
can be done in a way that avoids binning and the subsequent loss of
information.

Moreover, we have found that the analysis is very sensitive to the way
the volume derivative is calculated in equation \ref{eq:nIa}. Because
the SN sample is small, binning the data is not recommended, and all
the calculations should be done continuously. Binning can introduce a
relative overestimation of the volume derivative at low redshift,
effectively pushing the most probable time delays to higher values. As
a result, the lower limit of the Bayesian analysis can be
overestimated by more than $\sim 1$~Gyr and the peak of the Bayesian
probability distribution by $\sim 500$~Myr, according to our
calculations. This is consistent with the corrected results in
\citet{str04erratum}.

However, in order to get a continuous version of the control times we
have used an interpolation of the values calculated in S04 at redshift
intervals of $0.2$. Recalculating the control times with more redshift
resolution would be a better approach, but we have not tried it in
this analysis.

\subsubsection{Goodness of fit test (GoF)}

A maximum likelihood analysis must be accompanied by a GoF test to
check that the parametrized model has an appropriate form to start
with, and only then can the confidence intervals be trusted at
all. Accordingly, a $\chi^2$ test was used to check consistency
between the predicted and observed redshift histograms of SNe Ia in
S04. However, the $\chi^2$ test is not reliable when the number of
elements per bin is not greater than five in 80 per cent of the bins
\citep{WJ03}. Instead, we have used a Kolmogorov--Smirnov test (KS
test) as our goodness of fit test, which is the recommended test to
use when the sample size is small and because the analysis is done
continuously. Furthermore, selecting the best-fitting models with the
KS test can be used as an alternative parameter estimator.

\subsubsection{Confidence intervals} \label{sec:ConfInt}

In S04 the confidence intervals were obtained starting from the mode
(the maximum) of the Bayesian probabilities, partly because in the
original results the probabilities at low time delays were negligible.
However, because in some cases the probability distributions are
relatively flat, or not negligible at zero time delay, the definition
of the confidence intervals becomes important. In our calculations,
taking 95 per cent confidence intervals around the maximum vs. from the
median can make a difference of typically $\sim 500$~Myr.

One way to avoid this problem it to use the parameter region that is
not rejected by the GoF test with a certain confidence level. In this
approach, we obtain 95 and 68 per cent confidence intervals that are
unambiguously defined.

\subsubsection{Photometric redshifts}

Because the spectra of SNe are characterised by many blended lines
broadened by high velocity dispersion, SN redshifts are determined from
their host galaxies. Of the 25 SNe Ia in the sample, six have only
photometric redshifts, three of them in each field. We found that the
photometric redshift of SN 2003al, $0.91 \pm 0.2$, has a better
estimate in the public COMBO--17 catalogue \citep{wol04} of $0.82 \pm
0.04$. Additionally, in \citet{str05}, the photometric redshift of SN
2003lu, $0.11 \pm ^{0.13}_{0.11}$, has a better estimate of $0.14 \pm
0.01$.

\section{Results} \label{sec:results}

\subsection{Varying the SFH} \label{sec:resultsSFH}

If alternative SFHs are allowed, the Bayesian probabilities
associated with a given time-delay distribution have a wide range of
preferred time delays. The Bayesian probabilities and KS test
associated rejection probabilities show a preference for values
ranging from very long ($\sim 4$~Gyr) to very short, and even negative
($\sim -3$~Gyr) if the SFH peaks very late in time (see
Fig.~\ref{fig:probs}). The relation between the peak of the SFH and
the peak of the Bayesian probability distribution is, to zeroth order,
such that later peaked SFHs give shorter time delays.

Interestingly, inspection of Fig.~\ref{fig:probs} (upper-left panel)
shows two types of maxima in the Bayesian probabilities: one whose
position decreases with later peaked SFHs and another that is fixed at
approximately $\sim 3.5$~Gyr, even for different SFHs. The first peak
approximately reflects the time difference between the peaks of the
SFH and the SNR. The second peak reflects the relative absence of SNe
at high--z. Because no SNe were detected between the epoch of
reionisation and $z \sim 1.5$, or between $t \sim 0.5$~Gyr and $t \sim
4$~Gyr, the Bayesian analysis marginally favours models that do not
produce SNe in the first $\sim 3.5$~Gyr after the assumed epoch of
reionisation. The upper plot of Fig.~\ref{fig:SNeHisto} is
illustrative of this effect.

With the current data, it is the first peak which is statistically
dominant for all the SFHs, but this may change with deeper and wider
SN surveys in the future.

\subsection{Kolmogorov--Smirnov test}

With the KS test we find best-fitting mean time
delays and confidence intervals that are free from the problems
explained in Section \ref{sec:ConfInt}. The rejection probabilities
for the `fixed width Gaussian' and e-folding time-delay
distributions are shown in Fig.~\ref{fig:probs}.

We found that all the combinations of SFH and time-delay distributions
had an associated parameter region that is accepted by the KS test,
which validates the use of the Bayesian analysis. Additionally, the
parameter estimation seems robust in the sense that it gives results
that are consistent with what the Bayesian analysis shows. Moreover,
the addition of this test shows that the negative time-delay peak for
the H04 SFH is favoured over the long time-delay peak (see
Fig.~\ref{fig:probs}), which may be an indication that this SFH is not
compatible with the SN data.

\subsection{Confidence intervals}

\begin{figure}
\centering
\includegraphics[clip=true,width=0.8\hsize, keepaspectratio]{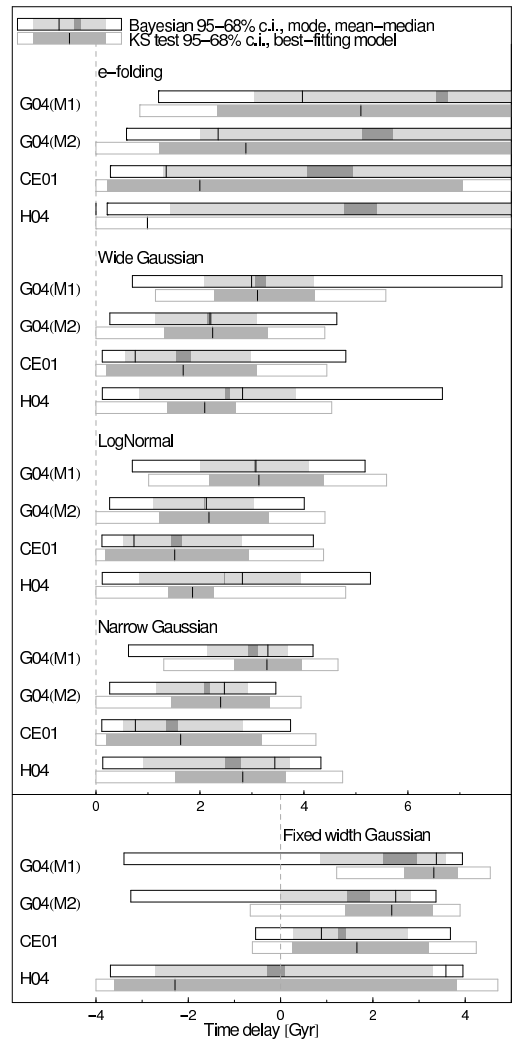}
\caption{Summary of the parameter estimation analysis using different
  SFHs and alternative time-delay distributions. The Bayesian
  probability 95 and 68 per cent confidence intervals are plotted with
  the position of the mode and the mean--median interval, in addition
  to the KS test rejection probabilities 95 and 68 per cent confidence
  intervals and the position of the best-fitting time delay. Note that
  later peaked SFHs show a preference for shorter time delays and that
  wider time-delay distributions allow for longer time delays. H04
  gives consistently poorer fits for positive time delays than for
  negative time delays as the use of the `fixed width Gaussian'
  distribution shows (see Fig.~\ref{fig:probs} and the discussion in
  Section \ref{sec:timedist} for the use of negative time delays).}
\label{fig:summaries}
\end{figure}

The confidence intervals were defined as the parameter region that
cannot be rejected with a certain confidence level based on the KS
test. We found that only the extinction corrected SFH from
\citet{gia04} has a 95 per cent confidence lower limit greater than zero,
i.e. around $1$~Gyr. All the alternative SFHs did not result in a
lower limit for the time delays greater than zero. A summary of the
confidence intervals obtained with the Gaussian and e-folding time
delay distributions is shown in Fig.~\ref{fig:summaries}.

\subsection{Varying the time-delay distribution}

The non-rejection regions of the five time-delay distributions tested
in this work can be grouped into three families of results: one family
associated with the `fixed width Gaussian' (that allows for negative
time delays) and `narrow Gaussian' distributions, another with the
`wide Gaussian' and log-normal distributions and one associated with
the e-folding distribution.

As a general rule, the narrower the test time-delay distribution, the
narrower the associated Bayesian probabilities. However, it is in the
long time-delay region where the changes are more noticeable, as can
be seen in Fig.~\ref{fig:summaries}. This is because the abrupt
transition that occurs in the SFH at the epoch of reionisation is
reflected in a less smooth SNR when narrower time-delay distributions
are assumed. Hence, the wider the time-delay distribution, the less
pronounced the second peak in the Bayesian probabilities (see Section
~\ref{sec:resultsSFH}) and the longer the time delays allowed.

\subsection{Theoretical time-delay distributions}

We performed KS tests of the theoretical time-delay distribution
varying the BPS parameters and the assumed SFHs. As a result, if the
extinction corrected SFH from G04 is assumed, the CO WD + MS -- SD
scenario alone has a 3 per cent probability of not being rejected and
the best-fitting models are obtained for a DD scenario which has a
very high mass transfer efficiency. Conversely, if any of the
theoretical models are assumed to be true, the best-fitting models
are, in almost all the combinations, obtained with the SFH from
\citet{CE01}. In Table \ref{tab:summary_models} we show the
non-rejection probabilities for the different combinations of SFH,
theoretical scenario and BPS parameters tested in this work. The BPS
parameters are: $\alpha_{\rm CE}$, the common--envelope ejection
efficiency efficiency as in \citet{HP04}; $\alpha_{\rm RLOF}$, the
Roche lobe overflow mass transfer efficiency, and $Z$, the
metallicity.

\begin{table}\centering
\caption{Summary of the KS non-rejection probabilities (per cent) for
  different combinations of SFH, theoretical time-delay distribution
  and BPS parameters. Unless stated otherwise, the standard parameters are
  $\alpha_{\rm CE} = 1.0$~, $\alpha_{\rm RLOF} = 1.0$~ and $Z =
  0.02$~.}
\begin{tabular}{@{}l ccc c ccc@{}}
\cmidrule{1-8}
 &
\multicolumn{3}{c}{SD scenario -- $\alpha_{\rm{CE}}:$} & &
\multicolumn{3}{c}{DD scenario -- $\alpha_{\rm{CE}}:$}  \\
\cmidrule{2-8}
SFH & 0.5 & 0.75 & 1.0 & & 0.5 & 0.75 & 1.0\\
\cmidrule{1-8}
 G04 (M1) & 3.1 & 3.0 & 3.1 & & 5.7 & 11.4 & 25.9 \\
 G04 (M2) & 12.2 & 11.7 & 12.4 & & 15.8 & 29.1 & 51.6 \\
 CE01 & 49.8 & 47.8 & 50.8 & & 48.1 & 69.2 & 85.4 \\
 H04 & 28.1 & 28.1 & 28.0 & & 25.7 & 24.2 & 21 \\
\cmidrule{1-8}
 &
\multicolumn{3}{c}{DD scenario -- $Z:$}  & &
\multicolumn{3}{c}{DD scenario -- $\alpha_{\rm{RLOF}}:$} \\
\cmidrule{2-8}
SFH & 0.001 & 0.004 & 0.02 & & 0.5 & 0.75 & 1.0 \\
\cmidrule{1-8}
 G04 (M1) & 11.0 & 13.2 & 25.9 & & 25.9 & 47.4 & 56.5 \\
 G04 (M2) & 30.1 & 31.2 & 51.6 & & 51.6 & 82.6 & 78.5 \\
 CE01 & 70.2 & 71.2 & 85.4 & & 85.4 & 80.7 & 62.5 \\
 H04 & 25.2 & 23.1 & 21.0 & & 21.0 & 18.1 & 17.3 \\
\cmidrule{1-8}
\end{tabular}
\label{tab:summary_models}
\end{table}

\begin{figure}
\centering
\vbox{
\includegraphics[clip=true,width= 0.4\hsize, keepaspectratio]{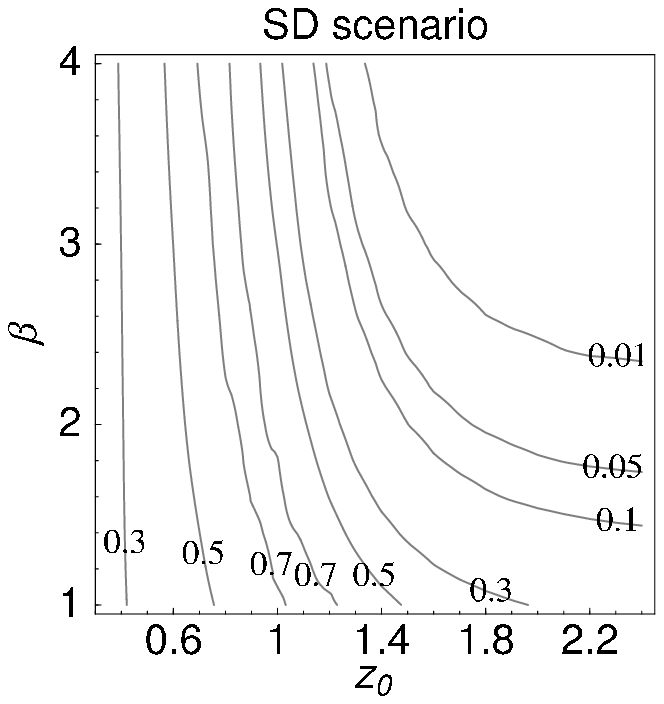}
\includegraphics[clip=true,width= 0.4\hsize, keepaspectratio]{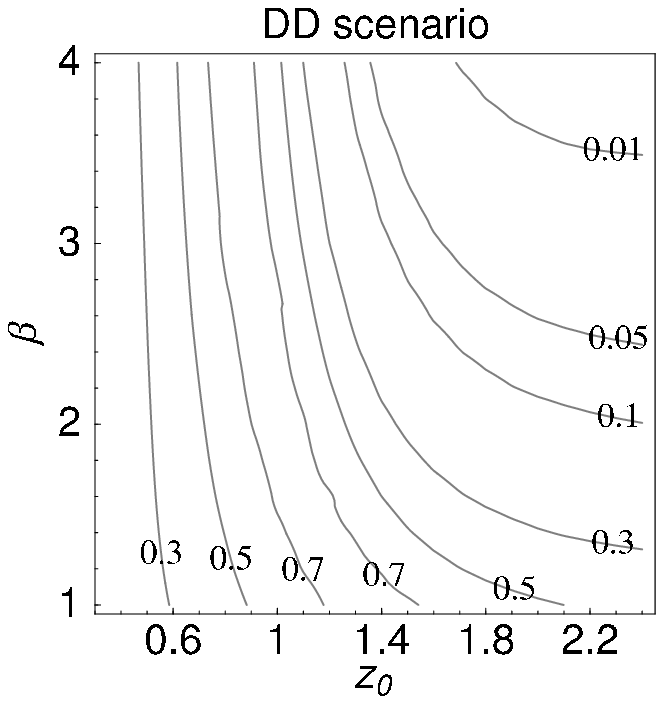}
}
\caption{Constraints on the SFH assuming the theoretical time delay
  distributions. We show the KS test marginalised probabilities of
  not-rejecting a model with parameters $z_0$ and $\beta$ (see
  equation~\ref{eq:SFRGY}). The probabilities are marginalised over
  $\alpha$ in the range -2 to 0. We have also marginalised over
  $\beta$ or $z_0$ and found that for the purpose of this work, the
  most important parameter is $z_0$, followed by $\beta$ and $\alpha$
  in order of importance.}
\label{fig:KS_z0beta}
\end{figure}

\subsection{Constraints on the SFH assuming the theoretical models}

In addition to the previous constraints on the time delay distribution
we followed the approach of \citet{GYM04} to constrain the SFH. The
method assumes a particular time delay distribution and a SFH that
consists of a broken power--law smoothly joined at the transition
redshift $z_0$, proportional to $\sim (1+z)^\alpha$ at high $z$ and to
$\sim (1+z)^\beta$ at low $z$, i.e.:

\begin{align}
 SFR(z) \propto \left\{ \left( \frac{1+z_0}{1+z} \right)^{5 \alpha} +
 \left ( \frac{1+z_0}{1+z} \right) ^{5 \beta} \right\} ^{-1/5}.
\label{eq:SFRGY}
\end{align}
For more details see \citet{GYM04}. We assumed the theoretical time
delay distributions and performed a KS test for different combinations
of power--law indices and transition redshift, $z_0$. We marginalised
the probabilities on $\alpha$ because it was found that the dependency
on this parameter is weak, which is expected because the SN sample
contains very few objects at high redshift. The resulting probability
distributions for different values of $z_0$ and $\beta$, assuming
either the SD or DD scenario is shown in
Fig.~\ref{fig:KS_z0beta}. Virtually all measurements of the SFH
suggest $SFR(z) \propto (1+z)^\beta$, with $\beta$ in the range
$\sim$~2 to 4 \citep{wil02, PG05}. For $\beta \gtrsim 2$ we find that
the peak of the SFH is most likely between $z \sim 0.7$ and
$1.2$. However, any rejection at 90 per cent confidence is only
possible for a peak location at $z \gtrsim 2$. The flatter the recent
decline in the SFH (lower $\beta$), the less clearly constrained is
the peak of the SFH.

\begin{figure}
\centering
\vbox{
\includegraphics[clip=true,width=0.7\hsize, keepaspectratio]{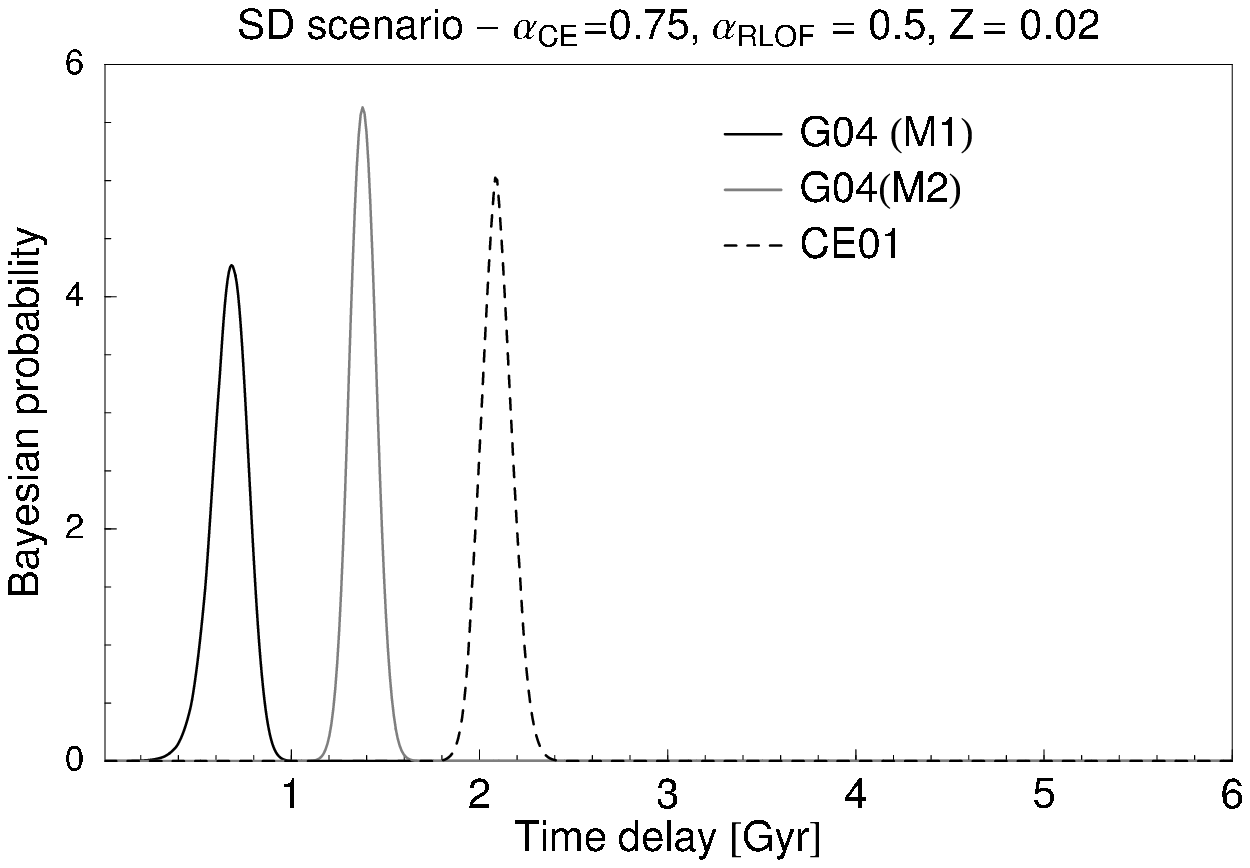}
}
\vbox{}
\vbox{
\includegraphics[clip=true,width=0.7\hsize, keepaspectratio]{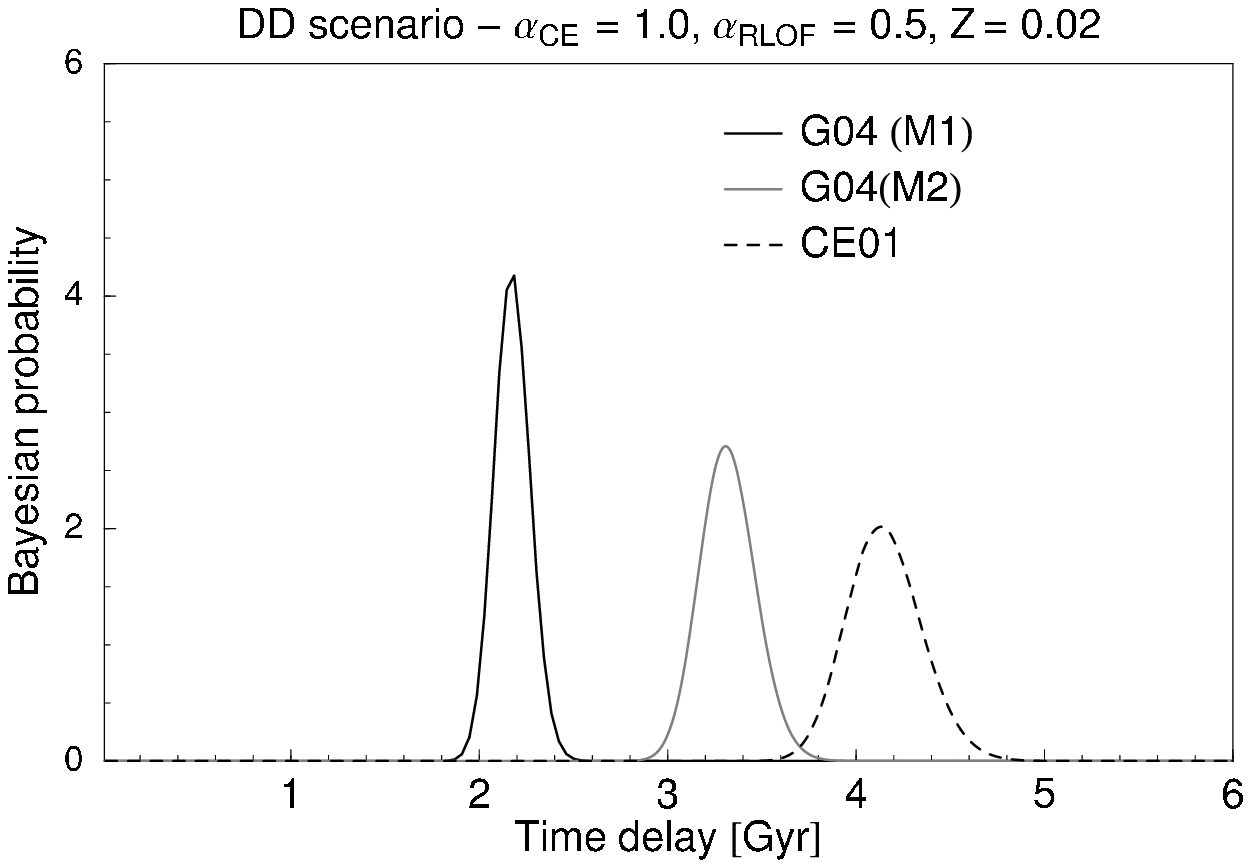}
}
\caption{Systematic errors associated with the choice of SFH. We
  performed Monte Carlo simulations assuming two different SFHs: the
  labelled SFH to derive the mock sample of SNe, but the G04 (M1) SFH
  to derive the Bayesian probabilities. We show the Bayesian
  probabilities derived from 10,000 mock SNe for one combination of
  BPS parameters in the SD and DD scenarios, respectively. The
  associated systematic error are of the order of $\sim 2$~Gyr.}
\label{fig:Monte Carlo}
\end{figure}

\begin{figure}
\centering
\includegraphics[clip=true,width=0.7\hsize, keepaspectratio]{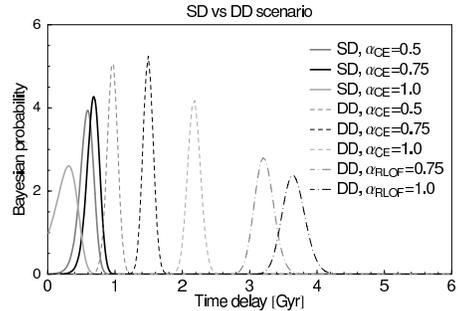}
\caption{Systematic errors associated with the choice of BPS
  parameters. The G04 (M1) SFH is assumed and the BPS parameters are
  changed. Because the time delay distribution are not fully sampled,
  the mean time delays are underestimated in some cases, but by less
  than $0.5$~Gyr. If some BPS parameters are changed, differences of
  up to $\sim 3-4$~Gyr are found for the \emph{mean} time delay of the
  DD scenario, while for the SD scenario they are of the order of
  $\sim 0.5$~Gyr.}
\label{fig:Monte Carlo2}
\end{figure}

\section{Monte Carlo simulations} \label{sec:Monte Carlo}

In this section we examine whether our approach is subject to systematic
biases due to the method itself and the choice of the SFH.

To assess the robustness of the method, we performed Monte Carlo
simulations for each progenitor scenario separately, where we took
large sets of simulated SNe (typically 10,000) to minimize statistical
errors. Because the DD scenario can extend to times comparable to the
age of the universe, its time delay distribution cannot be sampled
completely; we therefore expect that the recovered values will always
be biased towards shorter mean time delays. We found the following:
(1) such an expected bias is indeed present for the DD scenario, but
is always below $0.5$~Gyr; (2) for large sample sizes, the smallest
statistical errors are obtained using the e-folding time delay
distribution, independent of the theoretical scenario assumed.

To quantitatively estimate the systematic errors associated with the
choice of SFH, we proceeded in the following way: (1) in order to
minimize the statistical errors and study the systematics, we
performed Monte Carlo simulations with 10,000 mock SNe drawn from the
theoretical scenarios, using the e-folding time-delay distribution in
the analysis; (2) we produce a mock sample of SNe with a SFH that
differs from the one used in the Bayesian analysis; (3) a comparison
between the different recovered mean time delays then gives an
estimate of the systematic error due to the choice of SFH (see
Fig.~\ref{fig:Monte Carlo} and Fig.~\ref{fig:Monte Carlo2}). The result
is consistent with what can be concluded from
Section~\ref{sec:results}, i.e. that the bias on the mean time delay
can be of the order of $\sim 2$~Gyr, even ignoring the H04 SFH.

We then tested the robustness of the algorithm with small samples. In
order to do this, we performed Monte Carlo simulations with 10,000
different sets of 25 mock SNe drawn from the theoretical models. We
repeated the Bayesian and GoF analysis and found the following: (1)
when the theoretical time delays are relatively short and the SFH
peaks early in time, or when the time delays are long and the SFH
peaks late in time, the Bayesian probabilities tend to have very
positive or very negative skewness and, consequently, the mode of the
Bayesian probabilities either overestimates or underestimates the
theoretical mean time delays; (2) the distribution of non-rejection
probabilities using the KS test is flat, meaning that there is no
significant bias towards short or long time delays. Hence, we conclude
that the mode of the Bayesian probability is not a good estimator for
the mean time delays, while the KS test confidence intervals are
robust even with small samples, general results consistent with the
discussion in \citet{pre92}.


\begin{figure}
\centering 
\includegraphics[clip=true,width=0.9\hsize,keepaspectratio]{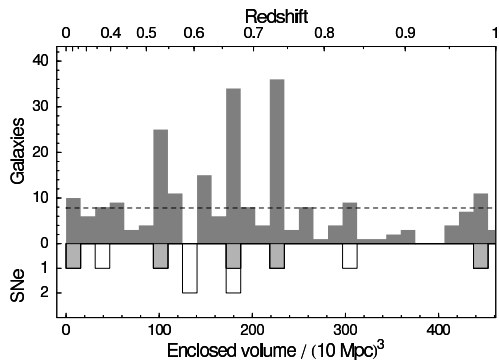}
\caption{Redshift histogram of a complete sample of galaxies in the
  GOODS--Southern field from the VVDS survey (gray), all the SNe with
  spectroscopic redshift in the field (white) and only the SNe Ia with
  spectroscopic redshift (light gray). Bins are chosen to show
  constant co-moving volume. The dashed line shows the average galaxy
  density (assuming no evolution). Several peaks are due to
  large-scale structure. Note that at higher redshifts the width of
  structures appears artificially enhanced by the distorted redshift
  scale.}
\label{fig:vvds_histo}
\end{figure}

\section{Discussion}  \label{sec:discussion}

\subsection{Large scale structure}

Large-scale structure (LSS) effects are important in small pencil beam
surveys when studying the time delay of SNe Ia, even when the SFH and
the SNR have been measured in the same field. Usually, the SFR is
measured as an `instantaneous' observed rate at a particular location
in space and time, where for this study a prediction of the SNR will
be needed for comparison. However, this prediction ought to be based
on the SFR in the same position in space, not only at the time when
the SNe are seen to explode, but at earlier times as well.

Since the latter figure can not be measured directly, we have to
obtain the SFR for earlier times by looking at a higher redshift,
i.e. at a different location in space. Cosmic variance and the
dependence of star formation on local environment will lead to a SFR
measurement  that is somewhat different from the past SFR of the
target location in the same field. This SFR variance will appear
whether the SFR is measured from the same fields or not, since the
locations in space where the SFR and SNR are measured will be unrelated. A
1~Gyr time delay already translates into proper distance differences
of $\sim 500$~Mpc at $z=1$.

So, if we do not know the individual star-formation histories of
galaxies in the supernova field, we are best advised to use our best
knowledge of the cosmic SFH, while LSS still leaves an imprint on the
observed SNR history. We investigate the large scale structure in the
CDFS where the spectroscopic redshift survey \emph{VIMOS VLT Deep
  Survey} \citep[][VVDS]{LF04} overlaps almost precisely with the
GOODS field and is complete to $M_{\rm{V}} < -19.5$ at $z<1$.  In
Fig.~\ref{fig:vvds_histo} we show a histogram of galaxy redshifts with
$M_{\rm{V}}<-19.5$ in bin sizes chosen to contain a constant co-moving
volume of $(25\ \rm{Mpc})^3$. A non-evolving and homogeneous galaxy
distribution should appear flat in this representation.  Well-known
over-densities or wall-like structures are clearly apparent,
especially in the redshift range from 0.5 to 0.75.  Two wall-like
structures near $z \approx 0.67$ and $z \approx 0.74$ are conspicuous
in the distribution and have been observed in both x-ray and optical
surveys \citep[see e.g.][for independent confirmation]{G03, wol04}.

Also shown is a histogram of SNe Ia with secure (spectroscopic)
redshift determinations. SNe with photo-z's only have been omitted due
to the large error in redshift which makes any association with
structures in the galaxy distribution difficult.

The average bin in the redshift range from 0 to 1 contains 7.8
galaxies, whereas the average SNe-weighted bin contains 16.7 galaxies
(2.14 times the normal average), a possible indication that the
inferred SNR is affected by LSS in this field and redshift range. In
order to understand how these numbers depend on the choice of our
bins, we have repeated the calculations for bin widths of $(\Delta
V)^{1/3} = (20,21.. 25)$~Mpc and found ratios of 2.52, 2.38, 2.15,
2.22, 2.00 and 2.14, respectively. Interestingly, the ratios that
result using only SNe Ia-weighted bins are even higher: 2.87, 2.68,
2.88, 3.29, 2.55 and 2.97.

The imprint of LSS in the SNR distribution could be corrected in the
analysis of time delays. If the underlying mass density on a given
direction $\hat n$ were of the form $\rho(\hat n,z)=\rho_0(z) (1 +
\delta(\hat n,z))$, the over-densities could be corrected by
multiplying the \emph{control times} by the same factor, $1 +
\delta(\hat n, z)$.  However, we do not correct for density
variations, because we lack a good determination of the LSS with
consistent quality in both fields.

\begin{figure*}
\centering
\includegraphics[clip=true,width=0.63\hsize, keepaspectratio]{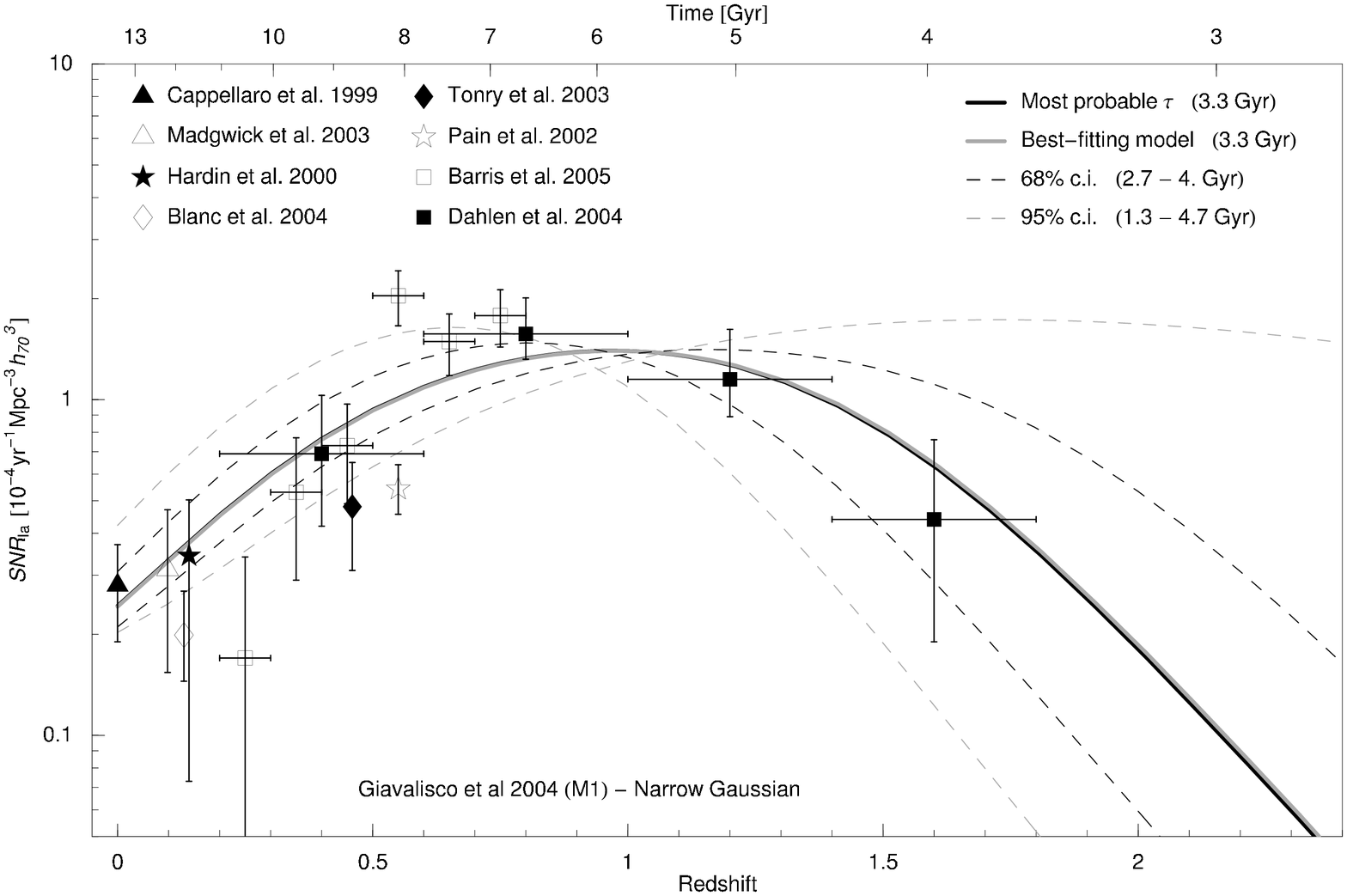}
\vbox{}
\includegraphics[clip=true,width=0.63\hsize, keepaspectratio]{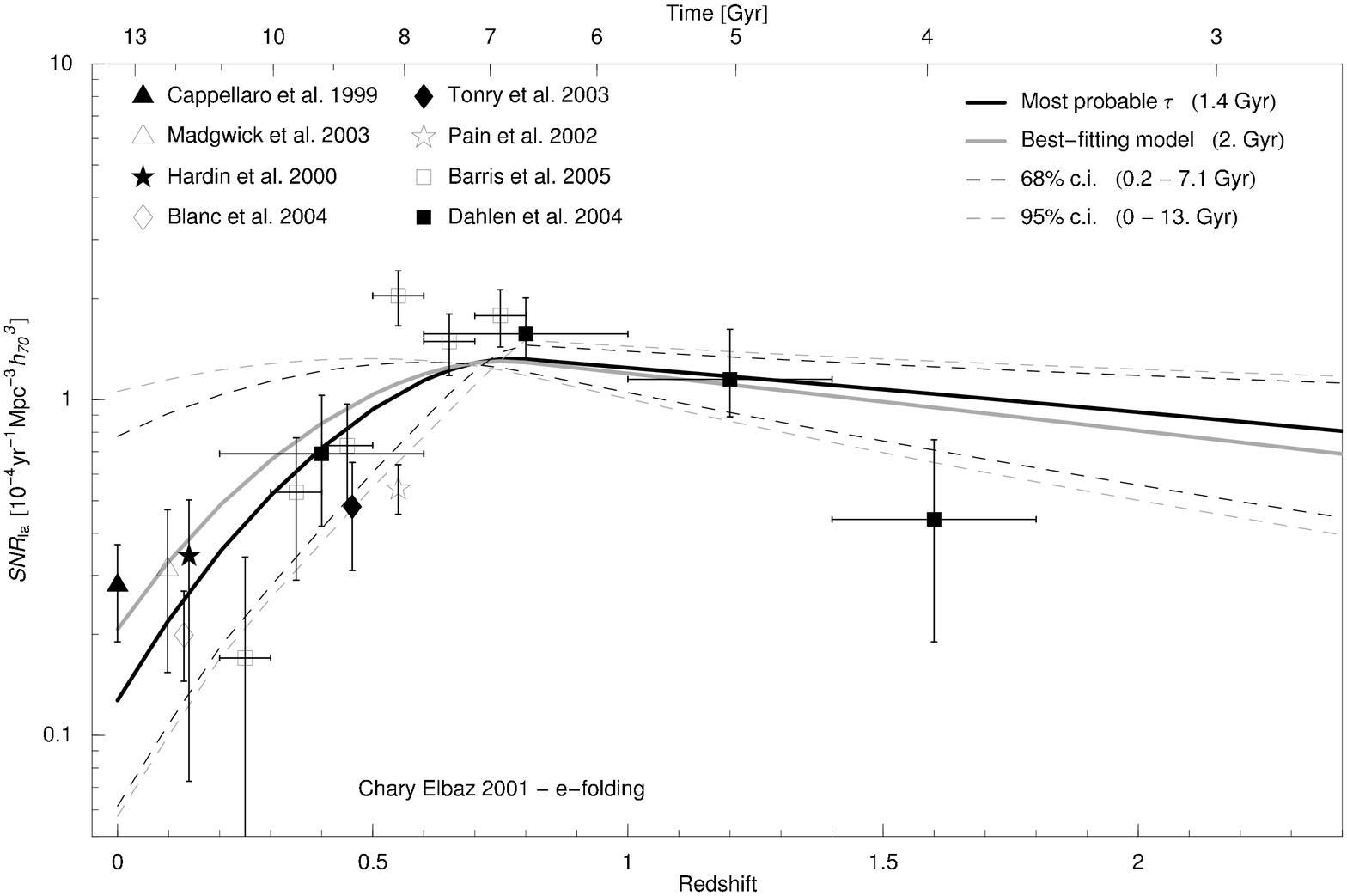}
\caption{Determinations of the SNR in the literature compared with the
  parametrized versions used in the analysis. The most probable $\tau$
  corresponds to the mode of the Bayesian probabilities and the
  best-fitting model corresponds to the model with the smallest KS
  test D value. The confidence intervals are based on the KS test,
  corresponding to the models where the non-rejection probabilities
  are 0.32 and 0.05, for the 68 and 95 per cent confidence intervals,
  respectively. The horizontal error bars are the bin sizes in
  \citet{dah04} and \citet{BT05}. Note that the highest redshift bin
  from \citet{dah04} is based on two detected SNe. The best-fitting
  models and confidence intervals are obtained using only the SN
  sample from S04.}
\label{fig:SNR}
\end{figure*}

\subsection{SN rates and efficiencies}

In the process of matching the observed number of SNe with a model, it
is possible to explicitly calculate the SN production efficiencies and
the supernova rate (SNR). Not only does the shape of the SNR history
constrain the progenitor models via the delay times, but also the SN
efficiency must be compatible with the model.

As a first consistency test, we compare the directly measured SNRs
with the parametrized estimates. In Fig.~\ref{fig:SNR} we show two
versions using the results of the parameter estimation with the G04
(M1) and CE01 SFHs. Clearly, the extinction corrected SFH from G04
with the `narrow Gaussian' distribution is consistent with long time
delays ($\sim 3.5$~Gyr), whereas the SFH from CE01 matches best with
an e-folding distribution of shorter time delays ($\sim 1.5$~Gyr).
The second alternative is only marginally favoured by the KS test, so
both possibilities seem equally plausible.

It is important to mention that in \citet{str05} a deeper SN search in
the smaller Ultra Deep Field and its parallel fields (UDF/P) resulted
in the detection of four additional SNe with $z<1.4$. The lack of
high-redshift SNe is one of the predictions of the S04 best-fitting
model and hence it supports the result. However, the authors also
considered the SFH from CE01, but without any time delay, and
concluded that it is not possible to rule out this SFH with more than
50 per cent confidence.

The SN efficiencies required to explain the observed SNR pose a problem 
for the theoretical SD scenario, which produces too few SNe. This is true 
for all SFHs and combination of parameters, and amounts to a shortfall of 
$4\times$ to $10\times$ depending on the assumed SFH. The DD scenario 
can reproduce the required efficiencies for some combinations of BPS
parameters. However, changes in the little constrained binary fraction 
and mass ratio distributions, or possibly in the initial mass function 
(IMF), may solve this problem in the future.

One additional challenge is that the time-delay distribution or the
supernova efficiencies may evolve with time: if the delay distribution
is made up of several components from different SN production
channels, their relative contribution could change with time; the
supernova efficiencies themselves may evolve with time, especially
considering that the accretion processes could depend on environmental
factors such as the metallicity \citep{kob98}. The relative absence of
SNe at high redshift could be a result of a metallicity effect, rather
than a reflection of the time delays. If this is truly the case, our
error bars on the time delays would be largely underestimated. The use
of compatible star formation and chemical enrichment histories will be
required to tackle this problem.

\subsection{\emph{Spitzer} SFH}

We have shown that it is crucial to determine the SFH that is to be
used in the analysis. The recent determination of the SFH from
\citet{PG05}, using three different extrapolations of the galaxy
luminosity function, was shown in Fig.~\ref{fig:loglog-SFH}. This SFH
was obtained by combining infrared \emph{Spitzer} observations with
optical--UV data of galaxies in the GOODS field. The resemblance
between this SFH and the CE01 best-fitting model justifies the use of
the latter.

Interestingly, the recent near--infrared and sub--millimetric
determination of the SFH from \citet{WCB05} peaks at $z \sim 1$,
similarly to what is obtained in CE01. These new results emphasize the
persistent uncertainties in our knowledge of the SFH and the problem
of extinction corrections in the optical.

\section{Conclusions} \label{sec:conclusions}

We have found that systematic errors associated with the use of
alternative star-formation histories are comparable if not larger than
the statistical errors reported in \citet{str04}.  The position of the
peak of the SFH was found to be the crucial parameter for the
recovered time delays: Later peaked SFHs result in lower time delays
and vice versa. Furthermore, the confidence intervals for the time
delays depend on the functional form of the delay distributions
assumed in the analysis. The use of wider time-delay distributions, in
particular the e-folding model, gives considerably longer upper limits
for the time delays.

For the data set under investigation, we found that the KS test is 
better suited to obtain confidence intervals than a Bayesian analysis. 
The KS test confidence intervals are unambiguously defined, and we have 
confirmed their validity using Monte Carlo simulations. In contrast, the 
skewness of the Bayesian probabilities and the small sample size result 
in somewhat arbitrary confidence intervals.

A KS test using the shape of the theoretical time-delay distributions
shows that the extinction corrected model from G04 is incompatible with 
the CO WD + MS -- SD scenario, although other SFHs are compatible. The 
DD scenario cannot be rejected at 95 per cent confidence with any combination
of SFH and time-delay distribution. Starting from theoretical time-delay 
distributions, they consistently favour the SFH from CE01.

If we wish to constrain the time-delay distribution and possibly discard 
progenitor scenarios from the redshift distribution of supernovae, it is
of foremost importance to determine the cosmic star-formation history 
more accurately. Otherwise, the uncertainty in the SFH will continue to 
limit the interpretation of SN data sets of any conceivable size.

Secondly, we would need a better understanding of our progenitor
models: if we ignore factors affecting efficiencies, their cosmic
evolution will make the delay time distributions evolve and produce a
situation where it is hard to disentangle these different effects
without qualitatively different observations.

Only after having solved these issues would deeper and wider surveys help to 
constrain the delay times and progenitor models by decreasing statistical
noise and reducing the influence of environmental cosmic variance on the
supernova samples. 

Different observations have already produced some evidence for shorter
time delays: Recent work by \citet{man05} showed that the most
efficient host galaxies for SNe Ia production among all galaxies are
irregulars, and \citet{dv05} showed that among elliptical galaxies it
is in particular the radio-loud galaxies, which are also believed to
be associated with recent star formation. Also, \citep{gar05} have
shown that SNe Ia of normal luminosity occur particularly in those
elliptical galaxies with quite substantial star-formation rates. Also,
SNe Ia in galaxy clusters seem to indicate time delays that are
shorter than 2~Gyr \citep{MGY04}.

Finally, it is clear that, if the SNe Ia phenomenon is composed of several
production channels, all conclusions we have drawn apply to the dominant 
channel. A generally less common channel with different characteristics
could again be the dominant channel in a subset of galaxies with older-age
or higher-metallicity stellar populations. It is already established that 
under-luminous 91bg-type SNe Ia preferably occur in non-star-forming hosts, 
such as ellipticals. It remains to be explored theoretically, whether the
CO WD + RG -- SD scenario could be related specifically to old populations.
We should also consider the possibility that we may not even have found 
the dominant progenitor channel for normal SNe Ia \citep{ham03, tou05}.

Even if the SFH peaks at redshift $\sim 1$ and the recovered time
delays are consequently low, the associated SNR (see
Fig.~\ref{fig:SNR}, lower panel) tends to be over-estimated in the
highest-z bin. This effect could be interpreted as the signature of a
long time delay component that does not contribute to the total SNR
when the universe is too young. However, because the highest-z bin
only contains two SNe, this does not lead to low non-rejection
probabilities. A study of bimodal time delay distributions could only
be done with this method if the uncertainties in the SFH were
significantly reduced and the metallicity cutoff on the efficiencies
was properly quantified.

If different channels produce SNe Ia from progenitors of distinctly
different age or metallicity, then an increase of the supernova sample
could greatly help the identification of the various plausible
progenitors.  However, such a data set would most be beneficial if it
is complemented with host galaxy characterisation 
\citep[see][]{vdb05} and spectra with better signal 
\citep[see][]{ben05}.

\section*{Acknowledgments}

We are indebted to Louis-G.Strolger for fruitful discussions and
providing us with the control times required in the analysis. We also
thank Guillaume Blanc, Ranga-Ram Chary, Ben Panther, Alan Heavens and
Pablo P\'erez-Gonz\'alez for discussions related to the SFH and Klaus
Meisenheimer and an anonymous referee for comments that significantly
improved the manuscript.  F.F. was supported by a Fundaci\'on Andes --
PPARC Gemini studentship. C.W. was supported by a PPARC Advanced
Fellowship.  This work was in part supported by a Royal Society
UK-China Joint Project Grant (Ph.P and Z.H.), the Chinese National
Science Foundation under Grant Nos. 10521001 and 10433030 (Z.H.) and a
European Research \& Training Network on Type Ia Supernovae
(HPRN-CT-20002-00303).



\begin{thebibliography}{}

\bibitem[\protect\citeauthoryear{Barris \& Tonry}{2005}]{BT05} Barris
  B.~J. \& Tonry J.~L., 2005, arXiv:astro-ph/0509655, \emph{accepted in
  ApJ}.
 
\bibitem[\protect\citeauthoryear{Benetti et al.} {2005}]{ben05}
Benetti S. et al., 2005, ApJ, 623, 1011

\bibitem[\protect\citeauthoryear{Blanc et al.}{2004}]{bla04} Blanc G.,
et al. 2004, A\&A, 423, 881

\bibitem[\protect\citeauthoryear{Cappellaro et al.} {1999}]{cap99}
Cappellaro, E., Evans, R., Turatto, M.\ 1999, A\&A, 351, 459
 
\bibitem[\protect\citeauthoryear{Chary \& Elbaz} {2001}]{CE01}
Chary R., Elbaz D., 2001, ApJ, 556, 562

\bibitem[\protect\citeauthoryear{Dahlen et al.} {2004}]{dah04} Dahlen
T. et al., 2004, ApJ, 613, 189

\bibitem[\protect\citeauthoryear{Della Valle et al.} {2005}]{dv05} 
Della Valle M., Panagia N., Padovani P., Cappellaro E., Mannucci F., 
Turatto M., 2005, ApJ,  629, 750 

\bibitem[\protect\citeauthoryear{Gal-Yam \& Maoz} {2004}]{GYM04}
Gal-Yam A., Maoz D., 2004, MNRAS, 347, 942

\bibitem[\protect\citeauthoryear{Gamezo Khokhlov \& Oran}
{2005}]{gam05} Gamezo V.~N., Khokhlov A.~M., Oran E.~S., 2005, ApJ,
623, 337

\bibitem[\protect\citeauthoryear{Garnavich \& Gallagher}
  {2005}]{gar05} Garnavich P.~M., Gallagher J., 2005,
  arXiv:astro-ph/0501065

\bibitem[\protect\citeauthoryear{Giavalisco et al.}  {2004}]{gia04}
Giavalisco M. et al., 2004, ApJ, 600, L103

\bibitem[\protect\citeauthoryear{Gilli et al.} {2003}]{G03} Gilli R.,
  et al.\ 2003, ApJ, 592, 721

\bibitem[\protect\citeauthoryear{Hardin et al.}{2000}]{har00} Hardin
D., et al.\ 2000, A\&A, 362, 419
 
\bibitem[\protect\citeauthoryear{Hachisu \& Nomoto} {1996}]{HKN96}
Hachisu I., Kato M., Nomoto K., 1996, ApJ, 470, L97
 
\bibitem[\protect\citeauthoryear{Hachisu et al.} {1999}]{HKN99} Hachisu I.,
Kato M., Nomoto K., 1999, ApJ, 522, 487

\bibitem[\protect\citeauthoryear{Hamuy} {2003}]{ham03} Hamuy M. et
al., 2003, Nat, 424, 651

\bibitem[\protect\citeauthoryear{Han \& Podsiadlowski} {2004}]{HP04}
Han Z., Podsiadlowski Ph., 2004, MNRAS, 350, 1301
 
\bibitem[\protect\citeauthoryear{Heavens et al.} {2004}]{hea04}
Heavens A., Panter B., Jimenez R., Dunlop J., 2004, Nat, 428, 625

\bibitem[\protect\citeauthoryear{Hillebrandt \& Niemeyer} {2000}]{HN00}
Hillebrandt W., Niemeyer J.~C., 2000, ARA\&A, 38, 191

\bibitem[\protect\citeauthoryear{Hopkins \& Beacom}{2006}]{HB06}
Hopkins A.~M., Beacom J.~F., 2006, arXiv:astro-ph/0601463,
\emph{submitted to ApJ}.

\bibitem[\protect\citeauthoryear{Iben \& Tutukov}{1984}]{IT84} Iben
I., Tutukov A.~V., 1984, ApJS, 54, 335

\bibitem[\protect\citeauthoryear{Kobayashi et al.} {1998}]{kob98}
Kobayashi C., Tsujimoto T., Nomoto K., Hachisu I., Kato, M., 1998,
ApJ, 503L, 155K
 
\bibitem[\protect\citeauthoryear{Langer et al.}
{2000}]{lan00} Langer N., Deutschmann A., Wellstein S., H{\"o}flich
P., 2000, A\&A, 362, 1046

\bibitem[\protect\citeauthoryear{Le F{\`e}vre et al.}{2004}]{LF04} Le
  F{\`e}vre O., et al.\ 2004, A\&A, 428, 1043

\bibitem[\protect\citeauthoryear{Li \& van den Heuvel} {1997}]{LV97}
Li X.-D., van den Heuvel E.~P.~J., 1997, A\&A, 322, L9
 
\bibitem[\protect\citeauthoryear{Mannucci et al.} {2005}]{man05}
Mannucci F., della Valle M., Panagia N., Cappellaro E., Cresci G.,
Maiolino R., Petrosian A., Turatto M., 2005, A\&A, 433, 807

\bibitem[\protect\citeauthoryear{Maoz \& Gal-Yam} {2004}]{MGY04} Maoz
  D., Gal-Yam A., 2004, MNRAS, 347, 951

\bibitem[\protect\citeauthoryear{Madgwick et al.}{2003}]{mad03}
Madgwick D.~S., Hewett P.~C., Mortlock D.~J., \& Wang L.\ 2003,
ApJ, 599, L33
  
\bibitem[\protect\citeauthoryear{Nomoto \& Iben} {1985}]{NI85} Nomoto
K., Iben I., 1985, ApJ, 297, 531
 
\bibitem[\protect\citeauthoryear{Nomoto \& Kondo} {1991}]{nom91}
Nomoto K., Kondo Y., 1991, ApJ, 367, L19

\bibitem[\protect\citeauthoryear{Pain et al.}{2002}]{pai02} Pain R.,
et al.\ 2002, ApJ, 577, 120

\bibitem[\protect\citeauthoryear{P{\'e}rez-Gonz{\'a}lez et al.} 
{2005}]{PG05} P{\'e}rez-Gonz{\'a}lez P.~G., Rieke G.~H., 
Egami E., et al., 2005, ApJ,  630, 82

\bibitem[\protect\citeauthoryear{Perlmutter et al.} {1999}]{per99}
Perlmutter S. et al., 1999, ApJ, 517, 565

\bibitem[\protect\citeauthoryear{Phillips} {1993}]{phi93} Phillips
M.~M., 1993, ApJ, 413, L105

\bibitem[\protect\citeauthoryear{Press et al.}{1992}]{pre92} Press
W.~H., Teukolsky S.~A., Vetterling W.~T., \& Flannery B.~P.\ 1992,
Numerical recipes in C: The art of scientific computing, Cambridge
University Press

\bibitem[\protect\citeauthoryear{Rappaport, Di Stefano, \& 
Smith}{1994}]{rap94} Rappaport S., Di Stefano R., Smith 
J.~D., 1994, ApJ, 426, 692 
 

\bibitem[\protect\citeauthoryear{Riess et al.} {1998}]{rie98} Riess
  A.~G. et al., 1998, AJ, 116, 1009

\bibitem[\protect\citeauthoryear{Riess et al.}  {2004}]{rie04} Riess
A.~G. et al., 2004, ApJ, 607, 665

\bibitem[\protect\citeauthoryear{R\"opke \& Hillebrandt}
 {2005}]{roe05} R\"opke F.~K., Hillebrandt W., 2005, A\&A, 431, 635

\bibitem[\protect\citeauthoryear{Saio \& Nomoto} {1985}]{SN85} Saio
H., Nomoto K., 1985, A\&A, 150, L21
 
\bibitem[\protect\citeauthoryear{Saio \& Nomoto} {1998}]{SN98} Saio
H., Nomoto K., 1998, ApJ, 500, 388
 
\bibitem[\protect\citeauthoryear{Shapiro \& Teukolsky} {1983}]{sha83}
Shapiro S.~L., Teukolsky S.~A., Black holes, white dwarfs and neutron
stars: The physics of compact objects
 
\bibitem[\protect\citeauthoryear{Strolger et al.} {2004}]{str04}
Strolger L.-G. et al., 2004, ApJ, 613, 200

\bibitem[\protect\citeauthoryear{Strolger et al.} {2005a}]{str05}
  Strolger L.-G. \& Riess A.~G., 2005, arXiv:astro-ph/0503093,
  \emph{submitted to AJ}.

\bibitem[\protect\citeauthoryear{Strolger et al.} {2005b}]{str04erratum}
 Strolger L.-G., et al. 2005, ApJ, 635, 1370

\bibitem[\protect\citeauthoryear{Timmes, Woosley, \& 
Taam}{1994}]{TWT94} Timmes F.~X., Woosley S.~E., Taam R.~E., 
1994, ApJ, 420, 348 

\bibitem[\protect\citeauthoryear{Tonry et al.}{2003}]{ton03} Tonry
J.~L., et al.\ 2003, ApJ, 594, 1
  
\bibitem[\protect\citeauthoryear{Tout} {2005}]{tou05} Tout C.~A.,
2005, ASPC, 330, 279

\bibitem[\protect\citeauthoryear{van den Heuvel et 
al.}{1992}]{vH92} van den Heuvel E.~P.~J., Bhattacharya D., 
Nomoto K., Rappaport S.~A., 1992, A\&A, 262, 97 
 
\bibitem[\protect\citeauthoryear{van den Bergh, Li \& Filippenko}
{2005}]{vdb05} van den Bergh S., Li W., Filippenko A.~V., 2005, PASP,
117, 773

\bibitem[\protect\citeauthoryear{Wall \& Jenkins} {2003}]{WJ03} Wall
  J.~V., Jenkins C.~R., 2003, Practical Statistics for Astronomers,
  Cambridge University Press

\bibitem[\protect\citeauthoryear{Wang, Cowie \& Barger} {2005}]{WCB05}
  Wang W.~H., Cowie L.~L. and Barger A.~J., 2005,
  arXiv:astro-ph/0512347, \emph{submitted to ApJ}.

\bibitem[\protect\citeauthoryear{Webbink}{1984}]{web84} Webbink R.~F.,
1984, ApJ, 277, 355

\bibitem[\protect\citeauthoryear{Wilson et al.} {2002}]{wil02} Wilson
  G., Cowie L.~L., Barger A.~J., Burke D.~J.\ 2002, AJ, 124, 1258
 
\bibitem[\protect\citeauthoryear{Wolf et al.}  {2004}]{wol04} Wolf C.,
Meisenheimer K., Kleinheinrich M., et al., 2004, A\&A, 421, 913

\end{thebibliography}
\end{document}